# Origin of Improved Photoelectrochemical Water Splitting in Mixed Perovskite Oxides

*Weiwei Li, Kai Jiang, Zhongguo Li, Shijing Gong, Robert L. Z. Hoye, Zhigao Hu, Yinglin Song, Chuanmu Tian, Jongkyoung Kim, Kelvin H. L. Zhang, Seungho Cho,\* and Judith L. MacManus-Driscoll\**

**Owing to the versatility in their chemical and physical properties, transition metal perovskite oxides have emerged as a new category of highly efficient photocatalysts for photoelectrochemical (PEC) water splitting. Here, to understand the underlying mechanism for the enhanced PEC water splitting in mixed perovskites, ideal epitaxial thin films of the BiFeO$_3$–SrTiO$_3$ system are explored. The electronic structure and carrier dynamics are determined from both experiment and density-functional theory calculations. The intrinsic phenomena are measured in this ideal system, contrasting to commonly studied polycrystalline solid solutions where extrinsic structural features obscure the intrinsic phenomena. It is determined that when SrTiO$_3$ is added to BiFeO$_3$ the conduction band minimum position is raised and an exponential tail of trap states from hybridized Ti 3d and Fe 3d orbitals emerges near the conduction band edge. The presence of these trap states strongly suppresses the fast electron–hole recombination and improves the photocurrent density in the visible-light region, up to 16× at 0 V$_{RHE}$ compared to the pure end member compositions. This work provides a new design approach for optimizing the PEC performance in mixed perovksite oxides.**

## 1. Introduction

Converting solar energy into chemical energy by photoelectrochemical (PEC) water splitting is a promising approach for sustainable energy production. In general, PEC involves absorption of the solar spectrum with the aid of semiconductor photoelectrodes to generate electron–hole pairs, followed by the oxidation and reduction of water to produce oxygen and hydrogen.[1] Enormous efforts have been invested in developing new photocatalytic materials with high photoconversion efficiencies based on wide absorption of the solar spectrum, efficient photogenerated charge carrier transfer, and high stability in an electrolyte solution. Owing to their versatility of chemical and physical properties, perovskites, a class of materials with highly symmetric close packed structures, have emerged as a new category of highly efficient photocatalysts without use of precious metals. The organic–inorganic lead halide perovskites, which are shown to be efficient photovoltaic materials with power conversion efficiencies above 20% developed within a few years, are a remarkable example.[2–6] However, their toxicity and instability issues pose significant challenges for applications in PEC cells. On the other hand, inorganic transition metal perovskite oxides, with the general formula of ABO$_3$, are very stable and offer great

Dr. W.-W. Li, Prof. J. L. MacManus-Driscoll
Department of Materials Science & Metallurgy
University of Cambridge
27 Charles Babbage Road, Cambridge CB3 0FS, United Kingdom
E-mail: jld35@cam.ac.uk

Dr. K. Jiang, Dr. S. Gong, Prof. Z. Hu
Key Laboratory of Polar Materials and Devices (MOE) and Technical Center for Multifunctional Magneto-Optical Spectroscopy (Shanghai)
Department of Electronic Engineering
East China Normal University
Shanghai 200241, China

Dr. Z. Li
College of Physics and Electronic Engineering and Jiangsu Laboratory of Advanced Functional Materials
Changshu Institute of Technology
Changshu 215500, China

Dr. R. L. Z. Hoye
Cavendish Laboratory
JJ Thomson Ave
University of Cambridge
Cambridge CB3 0HE, United Kingdom

Prof. Y. Song
Department of Physics
Harbin Institute of Technology
Harbin 150001, China

C. Tian, Prof. K. H. L. Zhang
College of Chemistry and Chemical Engineering
Xiamen University
Xiamen 361005, China

J. Kim, Prof. S. Cho
School of Materials Science and Engineering
Ulsan National Institute of Science and Technology (UNIST)
Ulsan 44919, Republic of Korea
E-mail: scho@unist.ac.kr

flexibility in functional properties. The bandgaps of perovskite oxides are typically derived from the B-site transition metal d-orbital in the conduction band and O 2p orbital in the valence band (VB). Most of the widely investigated perovskite oxides have a wide bandgap and are not suitable as photoactive layers. However, some single-component perovskite oxides such as $BiFeO_3$ (BFO) can absorb visible light, although they are prone to photocorrosion (the conduction band minimum and reduction potential lower than the water reduction potential, $H^+/H_2$, 0 V).[1,7] To overcome the limitations of single-component perovskite oxide photocatalysts, solid solutions consisting of two different perovskite materials have been actively investigated,[8] such as $Na_{0.5}La_{0.5}TiO_3$-$LaCrO_3$,[9] $AgNbO_3$-$SrTiO_3$,[10] $SrTiO_3$-$LaTiO_2N$,[11] and $BiFeO_3$–$SrTiO_3$.[12,13] By doping with 3d transition elements (e.g., Ti, Cr, and Fe), cations with $d^0$ or $d^{10}$ or $d^{10}s^2$ electronic configurations (e.g., Nb, Ag, and Bi), and nonmetal elements (e.g., N, C, and S), band edge positions can be effectively adjusted and thus solid solutions show significantly improved visible-light activity in comparison with those of their parent compounds.[14–17] Significant improvement in the electrochemical splitting of water has also been demonstrated by perovskite oxide solid solutions.[18–20] Prominent examples are the oxygen evolution reaction (OER) in $Ba_{0.5}Sr_{0.5}Co_{0.8}Fe_{0.2}O_{3-\delta}$,[21,22] the hydrogen evolution reaction (HER) in $Pr_{0.5}(Ba_{0.5}Sr_{0.5})_{0.5}Co_{0.8}Fe_{0.2}O_{3-\delta}$,[23] and the OER/HER in $SrNb_{0.1}Co_{0.7}Fe_{0.2}O_{3-\delta}$.[24]

In mixed-valent perovskites, the occupation of transition metal d orbitals and the hybridization between transition metal d and O p orbitals both play a critical role in tuning the energy band structure (e.g., bandgap, band edge position, and bandwidth). This highlights the importance of electronic structure in controlling the physical properties and catalytic activity.[9,10,17–19,25] In addition to electronic structure, the dynamics of the carriers (e.g., photoexcitation, relaxation of the excited state, and charge transfer/migration) is also of critical importance to understand the question of how multiple-electrons reaction in photocatalytic processes and to establish the mechanisms.[26] The dynamical process of hot electrons can be investigated by femtosecond time-resolved transient absorption (TA) spectroscopy, having been used in the well-studied binary water splitting systems of $TiO_2$ and $\alpha$-$Fe_2O_3$.[26–28]

To prevent the fast recombination of electron–hole pairs, the electronic structure of perovskite solid solutions needs to be linked to the dynamics of photoexcited carriers. Most previous solid solution photoelectrode work in perovskites has focused on polycrystalline solid solutions, but in such samples, the effects of their structural features (e.g., grain size, grain boundary, orientations, and specific surface areas) make the carrier dynamic process hard to probe. Hence, only very limited work has been undertaken on the simultaneous investigation of the electronic structure and dynamics of the carriers in perovskite solid solutions. Furthermore, the effect of defect states introduced by 3d transition element orbital hybridization on the migration of carriers and the performance of the PEC water splitting in the visible-light region has not been correlated to the carrier dynamics.

Perovskite BFO and $SrTiO_3$ (STO) are well-known photocatalytic materials owing to their relatively narrow bandgaps, high catalytic activity, and good chemical stability.[25,29–31] Our previous work reported that the PEC performance of epitaxial BFO–STO solid solution thin film can be carefully improved by changing the BFO–STO compositions.[12] However, the underlying origin of the improved PEC properties was not clearly elucidated. In this study, the correlation between electronic structure and carrier dynamics is studied in detail, which provides insight into the process of photoexcited carrier relaxation and recombination. From both experiments (UV-visible absorption spectra, Raman spectra, X-ray photoemission spectroscopy (XPS), and femtosecond time-resolved TA spectroscopy) and density-functional theory (DFT) calculations, we identified that, when $SrTiO_3$ is added to $BiFeO_3$, the conduction band minimum position is raised and an exponential tail of trap states from hybridized Ti 3d and Fe 3d orbitals emerges near the conduction band edge. These shallow trap states strongly prevent the fast recombination of electrons and holes in the mixed perovskite solid solution films and explain the origin of the enhanced photocurrent density in the visible-light region.

## 2. Results and Discussion

The quality and compositions of the BFO–STO solid solution films were investigated by X-ray diffraction (XRD), scanning electron microscope (SEM), and Raman spectra. Only (001) perovskite diffraction peaks were observed in the $\omega$-$2\theta$ scans (Figure S1, Supporting Information). There is no significant difference in microstructure in the top-view SEM images (Figure S2, Supporting Information). It is known that intensities of Raman spectra are directly proportional to stoichiometric ratios of thin films under the same measurement conditions. The intensities of the two phonon modes located at ≈ 310 and 679 cm$^{-1}$, which are associated with vibrations of the perovskite STO material, increase linearly with increasing STO molar ratio (Figure S3, Supporting Information). The results illustrate that the stoichiometric ratios are consistent with the designed compositions in the BFO–STO solid solution. However, it is found that the peak positions of the phonon modes related to the STO vibrations do not shift with different BFO compositions. Based on these results, it can be concluded that the deposited films are the high-quality epitaxial BFO–STO solid solution films.

To shed light on how the electronic structures of the BFO–STO solid solution films change with the composition, core-level photoemission spectra for all elements in the BFO–STO system were measured (**Figure 1**). The line shapes of most of the core-level spectra have almost no composition dependence. The 'contamination' signals associated with absorbed $H_2O$ or OH on the higher-binding-energy side of the O 1s peaks for the BFO–STO films are weak, indicating relatively clean surfaces (Figure 1e).[32] Compared with Fe 2p spectrum in the literature,[33] the Fe 2p spectra exhibit typical $Fe^{3+}$ features (Figure 1b).

In order to deduce the changes in the chemical potential, $\Delta\mu$, with increasing STO molar ratio, the core-level binding energy shift $\Delta E$ needs to be calculated:

$$\Delta E = -\Delta\mu + K\Delta Q - \Delta V_M + \Delta E_R \quad (1)$$

where $K$ is the coupling constant of the Coulomb interaction between the valence and the core electrons, $\Delta Q$ is the change in the number of valence electrons on the atom considered, $\Delta V_M$ is the change in the Madelung potential, and $\Delta E_R$ is the change in the extra-atomic relaxation energy caused by the screening of the core hole by metallic conduction electrons, or by polarization of the surrounding media for insulators.[32,34,35]

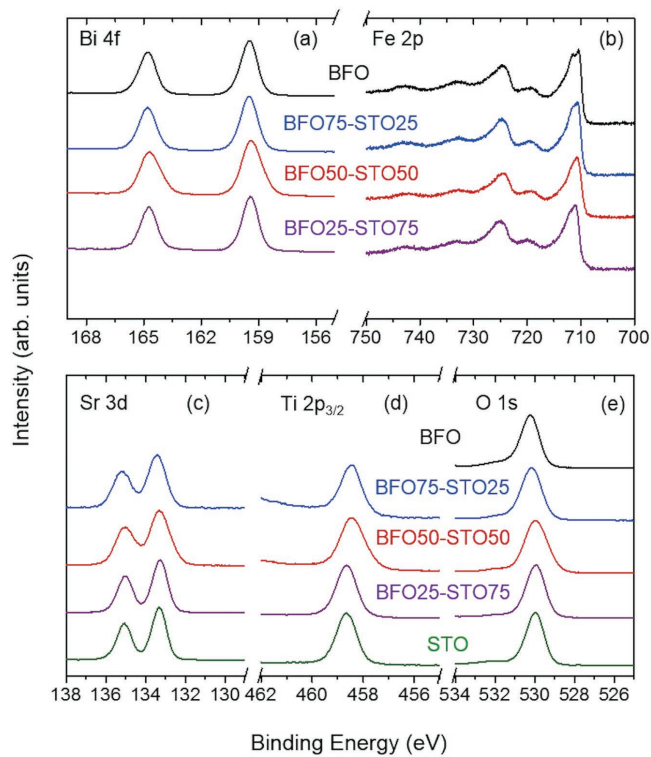

**Figure 1.** a) Bi 4f, b) Fe 2p, c) Sr 3d, d) Ti $2p_{3/2}$, and e) O 1s core-level spectra for BFO–STO solid solution films as a function of composition.

As seen in **Figure 2**a, all the core-level spectra, except for the Ti $2p_{3/2}$ core-level, shift toward lower binding energies with increasing STO molar ratio in the films. Returning to Figure 1d, the Ti $2p_{3/2}$ core-level moves in a different way from the other core-levels because the formal valence state of Ti changes with the composition of STO. It can be seen that the low-binding-energy side of the Ti $2p_{3/2}$ peak in BFO50-STO50 clearly exhibits an additional shoulder (Figure S4b, Supporting Information), indicating the appearance of $Ti^{3+}$, which might be caused by oxygen deficiency.[36,37] The similar shifts observed in Bi 4f, Fe 2p, Sr 3d, and O 1s imply that the change in the Madelung potential ($\Delta V_M$) can be neglected because it would shift the core-levels of cations and anion in different directions. Furthermore, the relative binding energy shift of the O 1s peak is almost twice those of Bi 4f, Fe 2p, and Sr 3d peaks, which may be due to both the BFO and STO contribution to the O 1s signal. There is no change in formal charge of Bi, Fe, Sr, and O for these compositions, resulting in $\Delta Q \approx 0$. Additionally, core-hole screening $\Delta E_R$ by conduction electrons can also be excluded from the main origin of the core-level shifts in transition metal oxides.[35,38] Therefore, we consider that the shifts measured from the Bi 4f, Fe 2p, and Sr 3d core-levels are largely due to the chemical potential shift $\Delta \mu$, and we take the average of the shifts of these three core-levels as a measure of $\Delta \mu$ in the BFO–STO solid solution films. Figure 2b shows the calculated $\Delta \mu$ values plotted as a function of percentage of STO in the BFO–STO solid solution films. A monotonic downward shift in chemical potential is observed with increasing STO molar ratio in the films and the shift has almost no change above the composition BFO25-STO75.

VB spectra of the BFO–STO solid solution films for the different molar ratios of STO are shown in **Figure 3**a. Also, the band structure and density of states (DOS) for the BFO, BFO50-STO50, and STO films in the energy regions near the Fermi level derived from the DFT band structure calculations are plotted in Figure 3b and Figure S5 in the Supporting Information. The calculated optical bandgaps for the BFO, BFO50-STO50, and STO films are 1.63, 1.84, and 2.23 eV, respectively. Considering the calculated value is underestimated by ≈20–30%,[39] the present DFT calculations are in good agreement with the experimental results (Figure S6, Supporting Information). We now return to the spectra for the pure BFO and the pure STO films as shown at the top and bottom of Figure 3a, respectively. The VB spectrum for pure BFO consists of four features labeled as A centered at ≈12 eV, B centered at ≈7.3 eV, C centered at ≈5.3 eV, and D centered at ≈2.5 eV. From the results of DFT calculations (Figure 3b and Figure S5, Supporting Information),[40] it is clear that feature A is dominated by Bi 6s states; feature B primarily originates from Fe 3d with a minor contribution from O 2p and Bi 6p; feature C arises from O 2p, having a minor admixture of Fe 3d and Bi 6p; and feature D is dominated by Fe 3d states with an admixture of O 2p character and a minor Bi 6s contribution, constituting the top portion of the VB. The VB spectrum of the pure STO film shows a feature E centered at ≈6.9 eV, which is dominated by the hybridization of Ti 3d and O 2p states with a minor Sr 3d contribution (Figure 3b and Figure S5, Supporting Information). A second feature F at ≈4.6 eV is primarily from O 2p states, but has a minor contribution from Ti 3d, forming the top of the VB.

It is noteworthy that the conduction band minimum of the BFO50-STO50 film moves toward a higher energy compared to the pure BFO film (Figure 3b). By combining the XPS-VB and DFT-DOS for the BFO50-STO50 film (Figure 3c), the schematic energy level diagram is proposed in Figure 3d. From Figure 2b, a monotonic downward shift in chemical potentials with increasing STO molar ratio can be understood by the replacement of heavy Fe with light Ti.[41] Figure 3a clearly shows that the intensity of feature D (Fe 3d states with an admixture of O 2p character and a minor Bi 6s contribution) decreases with increasing STO molar ratio. In this case, the O 2p level is then further from the Fe 3d orbital. As a consequence, the overlap of $t_{2g}$ and p orbitals is reduced because the oxygen p band is not strongly hybridized with the 3d band at the Fermi level.

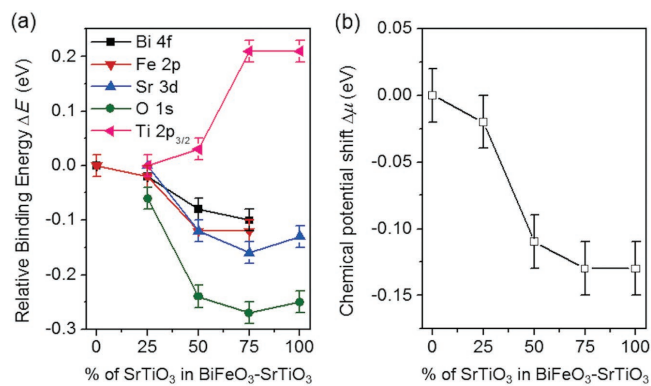

**Figure 2.** a) Binding energy shift of Bi 4f, Fe 2p, Sr 3d, Ti $2p_{3/2}$, and O 1s core-level, and b) average chemical potential shift ($\Delta \mu$) determined from the change of Bi 4f, Fe 2p, and Sr 3d binding energies as a function of STO composition.

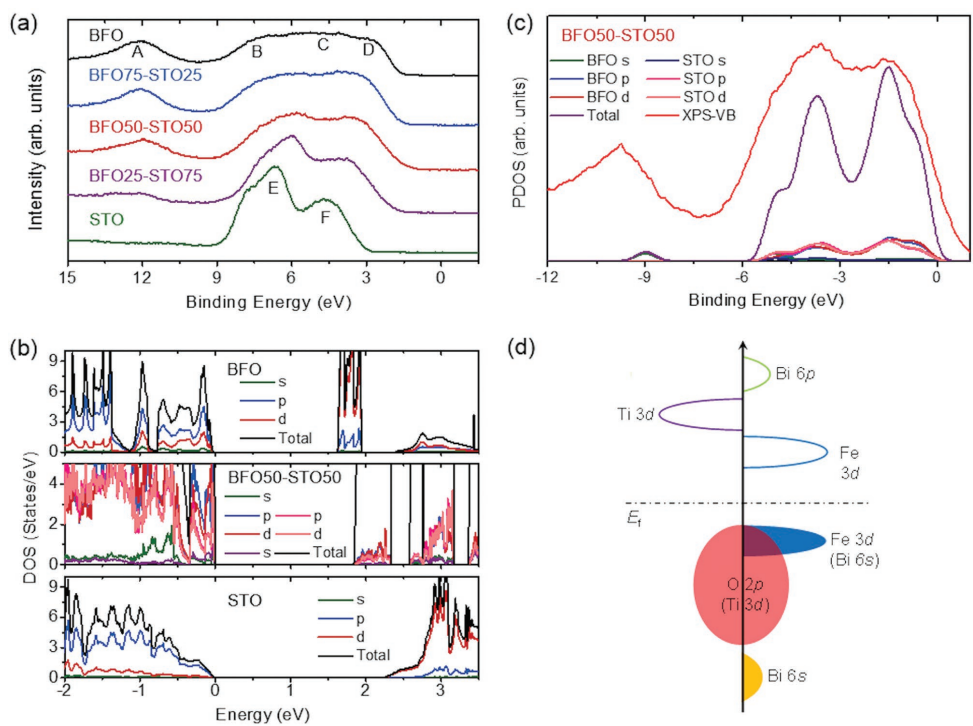

**Figure 3.** a) XPS-VB spectra for the BFO–STO solid solution film series. b) Total and partial DOS for pure BFO, BFO50-STO50, and pure STO from DFT calculation. c) Comparison of XPS-VB spectrum (only red curve) and calculated DOS, and d) schematic energy diagram for the BFO50-STO50 film.

TA spectroscopy was conducted to explore the relaxation dynamics of photoexcited carriers in the BFO–STO solid solution films. **Figures 4**a,b shows the typical TA decay curves for the BFO–STO solid solution films under 325 and 410 nm excitation, respectively. By considering the bandgaps of the BFO–STO solid solution films (Figure S5, Supporting Information) and the schematic energy level diagram (Figure 3d), the observed excitation at 325 nm (Figure 4a) is determined to originate from electrons transfer from the VB of O 2p to the conduction band of Ti 3d. The normalized TA recovery kinetics are similar for the different molar ratios of BFO and STO indicating similar recombination kinetics. Hence, changing the molar ratios does not significantly affect the electron dynamics within the first 2 ns following UV light excitation. This is ascribed to the low penetration depth of UV light in the films, as reported for other semiconductors (e.g., CdSe and TiO$_2$),[42,43] where the predominant carriers are generated within less than 50 nm from the surface.[44]

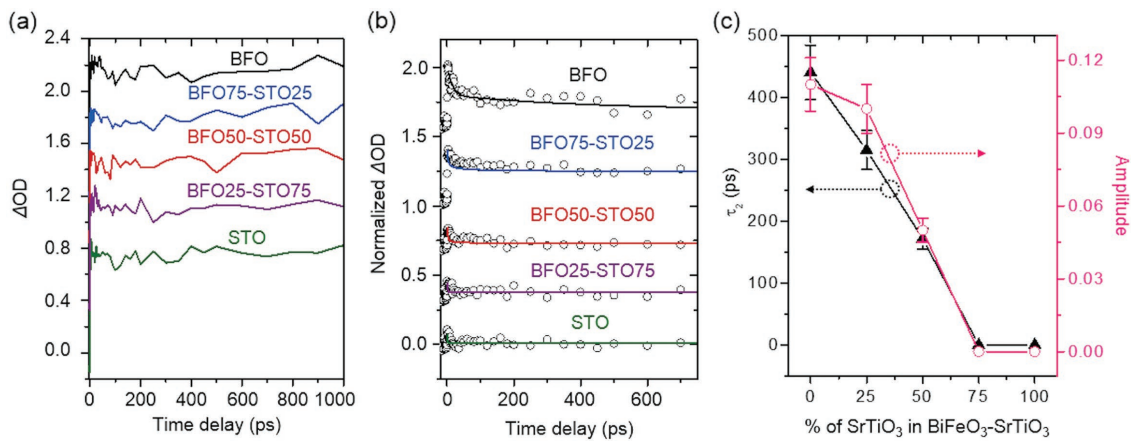

**Figure 4.** TA curves of the probe pulses through BFO–STO solid solution films under a) 325 nm and b) 410 nm excitation wavelength. The transient signal was probed at 532 nm. c) The dependence of carrier decay lifetime $\tau_2$ as a function of STO composition.

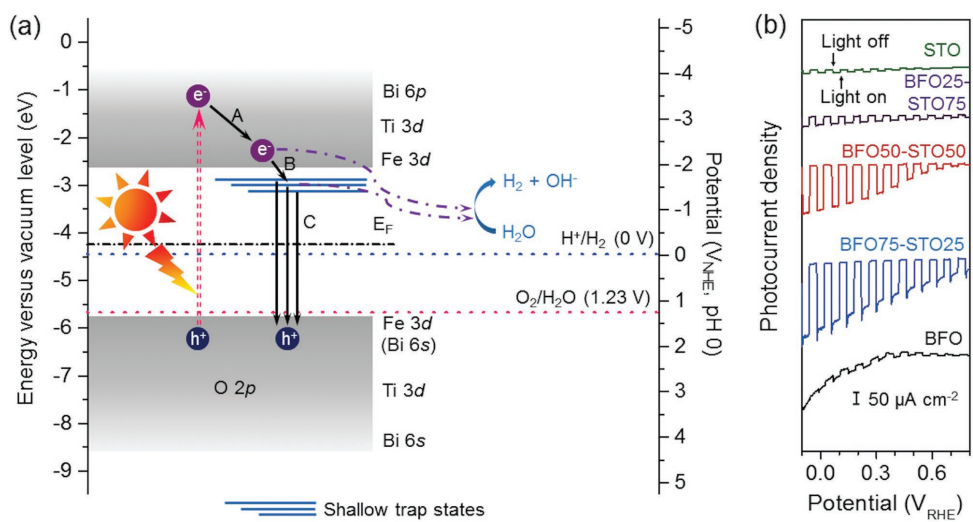

**Figure 5.** a) Schematic diagram of band positions relative to the vacuum level and NHE for the BFO50-STO50 film. A, B, and C represent the photoexcited carrier relaxation and recombination processes. b) Polarization curves of BFO, BFO75-STO25, BFO50-STO50, BFO25-STO75, and STO films. Reproduced with permission.[12] Copyright 2015, ACS.

Under 410 nm excitation, a different result is observed from the TA decay curves showing that the carrier relaxation dynamics can easily be tuned by varying the molar ratios between BFO and STO (Figure 4b). The TA data for the pure BFO film shows a maximum absorption intensity ≈ 2.33 eV (523 nm), which agrees well with the previously reported data.[45] Based on the electronic structures extracted from the XPS-VB and DFT calculations (Figure 3 and Figure S5, Supporting Information), the VB and conduction band edges relative to the vacuum level and normal hydrogen electrode (NHE) potential and the energy band structure of the BFO50-STO50 film are schematically illustrated in **Figure 5**a. This plot can be used to further understand the charge relaxation process. First, the photoexcited electrons should be transferred from the VB of O 2p to the conduction band of Bi 6p. Then these electrons release energy through electron–electron scattering or electron–phonon coupling within a time scale of ≈1 ps (procedure A in Figure 5a).[46–48] We note that the sub-ps decay process cannot be resolved in the TA measurements due to the pulse-width limitation of our TA system.

The ps to ns recovery kinetics can be fitted by a biexponential decay model[49]: $\Delta OD = A_1 e^{-t/\tau_1} + A_2 e^{-t/\tau_2}$. The experimental data of Figure 4b can be fitted very well using the solid lines by this decay model. All the films excited at 410 nm show fast decay lifetimes, $\tau_1$, on the order of 10 ps, which is related to the relaxation of photoexcited electrons near the conduction band edge to the exponential tails of shallow trap states generated by the hybridization between Ti 3d and Fe 3d states (procedure B in Figure 5a).[46]

In Figure 4c, it is seen that with increasing STO molar ratio the amplitude and lifetime $\tau_2$ of the slow decay component linearly decreases from 0.11 and 441 ps, respectively, for pure BFO. For pure STO, both these values are ≈0. The slow decay component can be attributed to recombination between electrons in the exponential tails states and holes in the VB (procedure C in Figure 5a). These results demonstrate that the relaxation dynamics and charge extraction of photogenerated carriers at 410 nm excitation strongly depend on the STO molar ratio.

As we discussed above, the conduction band minimum of the BFO50-STO50 film moves to a higher energy (Figure 3b), thus having a more negative reduction potential (Figure 5a). The TA results clearly demonstrate that the exponential tails of trap states induced by the hybridization between Ti 3d and Fe 3d orbitals result in efficient charge extraction which suppresses the rapid recombination of photogenerated carriers. This suppression enables the timescale for the HER to be strongly increased. This is consistent with recent works showing that the exponential tails of shallow trap states in binary $TiO_2$ polycrystalline samples play a key role in preventing the fast recombination of electrons and holes, thus enhancing the production of hydrogen.[50–52] The photocurrent density versus potential plots of the BFO–STO films are shown in Figure 5b. Compared to the pure end compositions (BFO and STO), the photocurrent density of the BFO75-STO25 and BFO50-STO50 films in the visible-light region is enhanced up to 16× at 0 $V_{RHE}$. The improved photocurrent and the production of $H_2$ (Figure S7, Supporting Information) of the BFO75-STO25 and BFO50-STO50 films can be explained by both the raised conduction band edge and the longer carrier lifetimes which result from the presence of an exponential tail of shallow trap states introduced by the hybridization of Ti 3d and Fe 3d orbitals. The incident photon-to-current efficiency (IPCE) of the BFO–STO solid solution films (Figure S8, Supporting Information) is higher than that of pure BFO and STO films in the visible-light region, in good agreement with the polarization curves shown in Figure 5b. Furthermore, the electrochemical impedance spectroscopy (EIS) results (Figure S9, Supporting Information) demonstrate that the fitted charge transfer resistance ($R_{ct}$) across the interface of electrode and electrolyte in the BFO–STO solid solution films is smaller than that of pure BFO and STO films and is reduced with the increase of BFO molar ratio. The reduction of $R_{ct}$ indicates an improved efficiency of photogenerated charge transport at the electrode/electrolyte interface. The EIS results agree well with PEC and IPCE performance.

## 3. Conclusion

In the mixed perovskite BiFeO$_3$–SrTiO$_3$ system, from both experiment and theory, the relation between electronic structure and carriers dynamics was studied in ideal epitaxial thin films. These films allowed intrinsic effects to be probed, constrasting to commonly polycrystalline samples where a wide range of extrinsic defects dominate the measurements. As well as increasing the bandgap and shifting the conduction band edge up, the addition of SrTiO$_3$ to BiFeO$_3$ leads to the hybridization of Ti 3d and Fe 3d orbitals. This produces an exponential tail of shallow trap states near the conduction band edge. At the same time, the recombination of electrons and holes is strongly suppressed, carrier lifeimes are increased, and the photocurrent density in the visible-light region is significantly increased. Our work is helpful to understand the mechanism of the enhanced performance in mixed perovskites for PEC water splitting and sets out a new design approach to improve PEC water splitting in perovskite solid solutions via hybridization of transition metal d orbitals.

## 4. Experimental Section

*Film Fabrication*: Epitaxial BFO, STO, and BFO–STO solid solution films (named as BFO100, BFO75-STO25, BFO50-STO50, BFO25-STO75, and STO100) were fabricated on (001)-oriented 0.5 wt% Nb-doped SrTiO$_3$ (primarily to prevent charging effects during XPS measurements) or STO substrates. For example, a solid solution film of BFO–STO with the molar percentage of 50:50 is denoted as BFO50-STO50. Details on sample preparations can be found elsewhere.[12]

*Characterization*: The crystalline nature of the films was investigated by XRD on a high-resolution X-ray diffractometer (Empyrean, PANalytical, The Netherlands) using Cu K$_\alpha$ radiation ($\lambda = 1.5405$ Å). XPS was measured by a monochromatic Al K$\alpha_1$ X-ray source ($h\nu = 1486.6$ eV) using a SPECS PHOIBOS 150 electron energy analyzer with a total energy resolution of 500 meV. The binding energy scale was calibrated using a polycrystalline Au foil placed in direct electrical contact with the film surface on the bench after deposition. The morphologies of the films were characterized by scanning electron microscope ((SEM); S-4800 Hitachi, Japan). The transmittance spectra were recorded on heating by a double beam near infrared-ultraviolet spectrometer (PerkinElmer UV/VIS LAMBDA 950) in the photon energy range of 190–2650 nm (0.5–6.5 eV) with an interval of 2 nm. Raman scattering experiments were carried out using a Jobin-Yvon LabRAM HR 800 UV micro-Raman spectrometer, excited by a 632.8 nm He–Ne laser and recorded in the frequency range of 50—1500 cm$^{-1}$ with a spectral resolution of 0.5 cm$^{-1}$. The wavelength-dependent TA measurements were performed using a standard femtosecond white-light pump-probe spectroscopy.[53] The light source was an optical parametric amplifier (Light Conversion ORPHEUS) pumped with a mode-locked Yb:KGW-based femtosecond laser (1.20 eV, 190 fs, 6 kHz). The pump wavelengths used in the present work were fixed at 325 and 410 nm. The white-light probe pulses were generated by focusing a 1.20 eV laser pulse onto a sapphire plate. To suppress many-body effects, the pump fluence was kept below 25 µJ cm$^{-2}$. Time resolution of the measurement system was ≈280 fs. All measurements were performed at room temperature. A photon energy 3.03 eV (410 nm) was chosen to selectively excite BFO, while both STO and BFO were excited under 3.82 eV (325 nm). The current–potential (*I*–*V*) curves of photoelecrochemical water reduction on BFO, STO, and BFO50-STO50 films on SrRuO$_3$-buffered STO substrates were obtained with a scan rate of 20 mV s$^{-1}$ in a phosphate solution (0.5 M, pH 12) using a platinum foil counter electrode, a Hg/HgO (1 M NaOH) reference electrode, and a Potentiostat/Galvanostat (CH Instruments, CHI604C). Prior to a measurement, the solution was purged with nitrogen for 30 min. The photocathodes were illuminated with AM 1.5 solar simulator (100 mW cm$^{-2}$, Newport Oriel 96 000) calibrated by a thermopile optical detector (Newport, Model 818P-010-12). The evolved amounts of H$_2$ were analyzed by a gas chromatograph (HP5890) with a thermal conductivity detector and a molecular sieve 5-A column. Incident photon-to-current efficiency (IPCE) was measured using the 150 W Xe lamp and a monochromator (Jobin Yvon Flurolog3, bandpass 5 nm). The EIS measurements were performed at 0.23 V versus RHE in 0.5 M potassium phosphate buffer (pH 12) under 1 sun illumination.

*DFT Calculation*: In order to investigate the origin of the electronic transitions in the BFO–STO solid solutions, first-principles calculations within DFT as implemented in the Vienna Ab-initio Simulation Package was performed.[54] The exchange-correlation potential was treated in the local density approximation. The surface Brillouin zone was sampled with *k*-point 15 × 15 × 1 meshes, and the energy cutoff was set to 400 eV for the plane-wave expansion of the projector-augmented waves in the self-consistent calculations. The convergence of the total energy was checked by changing the number of sampling *k*-points, energy cutoff, and the thickness of the vacuum space. For geometry optimization, all the internal coordinates were relaxed until the Hellmann–Feynman forces were less than 1 meV Å$^{-1}$. To eliminate the interaction between adjacent slabs, a large enough vacuum thickness ≈20 Å along the *z* axis was adopted.

## Supporting Information

Supporting Information is available from the Wiley Online Library or from the author.

## Acknowledgements

W.L., K.J., and Z.L. contributed equally to this work. W.L. and J.L.M.-D. acknowledge support from EPSRC grant EP/L011700/1, EP/N004272/1, and the Isaac Newton Trust (Minute 13.38(k)). K.J. acknowledges the sponsorship by Shanghai Sailing Program (Grant No. 18YF1407200). Z.L. acknowledges the financial support from the National Natural Science Foundation of China (11704048). R.L.Z.H. acknowledges support from Magdalene College, Cambridge. K.H.L.Z. acknowledges financial support from the Thousand Young Talents Program of China. J.K. and S.C. acknowledge the 2018 Research Fund (1.180061.01) of UNIST (Ulsan National Institute of Science & Technology) and the National Research Foundation of Korea (NRF) grant funded by the Korea government (MSIP; Ministry of Science, ICT & Future Planning) (No. NRF-2017R1C1B5075626 and NRF-2018R1C6002342).

## Conflict of Interest

The authors declare no conflict of interest.

[1] S. Chen, L. W. Wang, *Chem. Mater.* **2012**, *24*, 3659.
[2] M. Liu, M. B. Johnston, H. J. Snaith, *Nature* **2013**, *501*, 395.

[3] G. Xing, N. Mathews, S. Sun, S. S. Lim, Y. M. Lam, M. Grätzel, S. Mhaisalkar, T. C. Sum, *Science* **2013**, *342*, 344.

[4] S. D. Stranks, G. E. Eperon, G. Grancini, C. Menelaou, M. J. P. Alcocer, T. Leijtens, L. M. Herz, A. Petrozza, H. J. Snaith, *Science* **2013**, *342*, 341.

[5] C. Wehrenfennig, G. E. Eperon, M. B. Johnston, H. J. Snaith, L. M. Herz, *Adv. Mater.* **2014**, *26*, 1584.

[6] N. J. Jeon, J. H. Noh, W. S. Yang, Y. C. Kim, S. Ryu, J. Seo, S. I. Seok, *Nature* **2015**, *517*, 476.

[7] L. Guo, J. Luo, T. He, S. Wei, S. Li, *arXiv*: 1801.01334.

[8] P. Kanhere, Z. Chen, *Molecules* **2014**, *19*, 19995.

[9] J. Shi, J. Ye, Z. Zhou, M. Ling, L. Gao, *Chem. - Eur. J.* **2011**, *17*, 7858.

[10] D. Wang, T. Kako, J. Ye, *J. Am. Chem. Soc.* **2008**, *130*, 2724.

[11] W. Luo, Z. Li, X. Jiang, T. Yu, L. Liu, X. Chen, J. Ye, Z. Zou, *Phys. Chem. Chem. Phys.* **2008**, *10*, 6717.

[12] S. Cho, J. W. Jang, W. Zhang, A. Suwardi, H. Wang, D. Wang, J. L. MacManus-Driscoll, *Chem. Mater.* **2015**, *27*, 6635.

[13] L. Lu, M. Lv, D. Wang, G. Liu, X. Xu, *Appl. Catal., B* **2017**, *200*, 412.

[14] W. Wang, M. O. Tadé, Z. P. Shao, *Chem. Soc. Rev.* **2015**, *44*, 5371.

[15] W. Wang, M. O. Tadé, Z. P. Shao, *Prog. Mater. Sci.* **2018**, *92*, 33.

[16] G. Q. Zhang, S. R. Sun, W. S. Jiang, X. Miao, Z. Zhao, X. Y. Zhang, D. Qu, D. Y. Zhang, D. B. Li, Z. C. Sun, *Adv. Energy Mater.* **2017**, *7*, 1600932.

[17] G. Zhang, G. Liu, L. Z. Wang, J. T. S. Irvine, *Chem. Soc. Rev.* **2016**, *45*, 5951.

[18] J. Suntivich, H. A. Gasteiger, N. Yabuuchi, H. Nakanishi, J. B. Goodenough, Y. Shao-Horn, *Nat. Chem.* **2011**, *3*, 546.

[19] A. Vojvodic, J. K. Nørskov, *Science* **2011**, *334*, 1355.

[20] Y. Zhu, W. Zhou, Z. Shao, *Small* **2017**, *13*, 1603793.

[21] J. Suntivich, K. J. May, H. A. Gasteiger, J. B. Goodenough, Y. Shao-Horn, *Science* **2011**, *334*, 1383.

[22] J. I. Jung, H. Y. Jeong, M. G. Kim, G. Nam, J. Park, J. Cho, *Adv. Mater.* **2015**, *27*, 266.

[23] X. Xu, Y. Chen, W. Zhou, Z. Zhu, C. Su, M. Liu, Z. Shao, *Adv. Mater.* **2016**, *28*, 6442.

[24] Y. Zhu, W. Zhou, Y. Zhong, Y. Bu, X. Chen, Q. Zhong, M. Liu, Z. Shao, *Adv. Energy Mater.* **2017**, *7*, 1602122.

[25] S. Kawasaki, K. Akagi, K. Nakatsuji, S. Yamamoto, I. Matsuda, Y. Harada, J. Yoshinobu, F. Komori, R. Takahashi, M. Lippmaa, C. Sakai, *J. Phys. Chem. C* **2012**, *116*, 24445.

[26] J. B. Baxter, C. Richter, C. A. Schmuttenmaer, *Annu. Rev. Phys. Chem.* **2014**, *65*, 423.

[27] S. R. Pendlebury, M. Barroso, A. J. Cowan, K. Sivula, J. W. Tang, M. Grätzel, D. Klug, J. R. Durrant, *Chem. Commun.* **2011**, *47*, 716.

[28] X. L. Wang, A. Kafizas, X. Li, S. J. A. Moniz, P. J. T. Reardon, J. W. Tang, I. P. Parkin, J. R. Durrant, *J. Phys. Chem. C* **2015**, *119*, 10439.

[29] F. T. Wagner, G. A. Somorjai, *Nature* **1980**, *285*, 559.

[30] F. Gao, X. Y. Chen, K. B. Yin, S. Dong, Z. F. Ren, F. Yuan, T. Yu, Z. Zou, J. M. Liu, *Adv. Mater.* **2007**, *19*, 2889.

[31] K. H. L. Zhang, R. Wu, F. Z. Tang, W.-W. Li, F. E. Oropeza, L. Qiao, V. K. Lazarov, Y. G. Du, D. J. Payne, J. L. Driscoll, M. G. Blamire, *ACS Appl. Mater. Interfaces* **2017**, *9*, 26549.

[32] K. H. L. Zhang, Y. Du, P. V. Sushko, M. E. Bowden, V. Shutthanandan, S. Sallis, L. F. J. Piper, S. A. Chambers, *Phys. Rev. B* **2015**, *91*, 155129.

[33] M. Descostes, F. Mercier, N. Thromat, C. Beaucaire, *Appl. Surf. Sci.* **2000**, *165*, 288.

[34] S. Hüfner, *Photoelectron Spectroscopy*, Springer-Verlag, Berlin **2003**.

[35] A. Fujimori, A. Ino, J. Matsuno, T. Yoshida, K. Tanaka, T. Mizokawa, *J. Electron Spectrosc. Relat. Phenom.* **2002**, *124*, 127.

[36] W.-W. Li, R. Zhao, L. Wang, R. J. Tang, Y. Y. Zhu, J. H. Lee, H. X. Cao, T. Y. Cai, H. Z. Guo, C. Wang, L. S. Ling, L. Pi, K. J. Jin, Y. H. Zhang, H. Y. Wang, Y. Q. Wang, S. Ju, H. Yang, *Sci. Rep.* **2013**, *3*, 2618.

[37] W.-W. Li, Q. He, L. Wang, H. Z. Zeng, J. Bowlan, L. S. Ling, D. A. Yarotski, W. R. Zhang, R. Zhao, J. H. Dai, J. X. Gu, S. P. Shen, H. Z. Guo, L. Pi, H. Wang, Y. Q. Wang, I. A. Velasco-Davalos, Y. J. Wu, Z. J. Hu, B. Chen, R.-W. Li, Y. Sun, K. J. Jin, Y. H. Zhang, H.-T. Chen, S. Ju, A. Ruediger, D. N. Shi, A. Y. Borisevich, H. Yang, *Phys. Rev. B* **2017**, *96*, 115105.

[38] A. Ino, T. Mizokawa, A. Fujimori, K. Tamasaku, H. Eisaki, S. Uchida, T. Kimura, T. Sasagawa, K. Kishio, *Phys. Rev. Lett.* **1997**, *79*, 2101.

[39] A. Cohen, P. Mori-Sanchez, W. Yang, *Science* **2008**, *321*, 729.

[40] J. B. Neaton, C. Ederer, U. V. Waghmare, N. A. Spaldin, K. M. Rabe, *Phys. Rev. B* **2005**, *71*, 014113.

[41] M. Imada, A. Fujimori, Y. Tokura, *Rev. Mod. Phys.* **1998**, *70*, 1039.

[42] E. Busby, N. C. Anderson, J. S. Owen, M. Y. Sfeir, *J. Phys. Chem. C* **2015**, *119*, 27797.

[43] A. Kafizas, X. Wang, S. R. Pendlebury, P. Barnes, M. Ling, C. Sotelo-Vazquez, R. Quesada-Cabrera, C. Li, I. P. Parkin, J. R. Durrant, *J. Phys. Chem. A* **2016**, *120*, 715.

[44] Y. Yamada, H. K. Sato, Y. Hikita, H. Y. Hwang, Y. Kanemitsu, *Appl. Phys. Lett.* **2014**, *104*, 151907.

[45] Y. Yamada, T. Nakamura, S. Yasui, H. Funakubo, Y. Kanemitsu, *Phys. Rev. B* **2014**, *89*, 035133.

[46] L. T. Quynh, C. N. Van, Y. Bitla, J.-W. Chen, T. H. Do, W.-Y. Tzeng, S.-C. Liao, K.-A. Tsai, Y.-C. Chen, C.-L. Wu, C.-H. Lai, C.-W. Luo, Y.-J. Hsu, Y.-H. Chu, *Adv. Energy Mater.* **2016**, *6*, 1600686.

[47] Y. M. Sheu, S. A. Trugman, Y.-S. Park, S. Lee, H. T. Yi, S.-W. Cheong, Q. X. Jia, A. J. Taylor, R. P. Prasankumar, *Appl. Phys. Lett.* **2012**, *100*, 242904.

[48] R. V. Pisarev, A. S. Moskvin, A. M. Kalashnikova, Th. Rasing, *Phys. Rev. B* **2009**, *79*, 235128.

[49] M. Cargnello, T. Montini, S. Y. Smolin, J. B. Priebe, J. J. D. Jaen, V. V. T. Doan-Nguyen, I. S. McKay, J. A. Schwalbe, M.-M. Pohl, T. R. Gordon, Y. Lu, J. B. Baxter, A. Bruckner, P. Fornasiero, C. B. Murray, *Proc. Natl. Acad. Sci. U. S. A.* **2016**, *113*, 3966.

[50] X. Chen, L. Liu, P. Y. Yu, S. S. Mao, *Science* **2011**, *331*, 746.

[51] F. Nunzi, E. Mosconi, L. Storchi, E. Ronca, A. Selloni, M. Gratzel, F. De Angelis, *Energy Environ. Sci.* **2013**, *6*, 1221.

[52] R. Li, Y. Weng, X. Zhou, X. Wang, Y. Mi, R. Chong, H. Han, C. Li, *Energy Environ. Sci.* **2015**, *8*, 2377.

[53] Z. G. Li, R. Zhao, W.-W. Li, H. Wang, H. Yang, Y. L. Song, *Appl. Phys. Lett.* **2014**, *105*, 162904.

[54] G. Kresse, D. Joubert, *Phys. Rev. B* **1999**, *59*, 1758.